# Recent Developments in China-U.S. Cooperation in Science

*Minerva* (forthcoming)

Caroline S. Wagner[1], Lutz Bornmann[2], Loet Leydesdorff[3]

**Abstract**

China's remarkable gains in science over the past 25 years have been well documented (e.g., Jin and Rousseau, 2005a; Zhou and Leydesdorff, 2006; Shelton & Foland, 2009) but it is less well known that China and the United States have become each other's top collaborating country. Science and technology has been a primary vehicle for growing the bilateral relationship between China and the United States since the opening of relations between the two countries in the late 1970s. During the 2000s, the scientific relationship between China and the United States--as measured in coauthored papers--showed significant growth. Chinese scientists claim first authorship much more frequently than U.S. counterparts by the end of the decade. The sustained rate of increase of collaboration with one other country is unprecedented on the U.S. side. Even growth in relations with eastern European nations does not match the growth in the relationship between China and the United States. Both countries can benefit from the relationship, but for the U.S., greater benefit would come from a more targeted strategy.

[1] Milton & Roslyn Wolf Chair in International Affairs, John Glenn School of Public Affairs, Battelle Center for Science and Technology Policy, The Ohio State University, Columbus, Ohio, USA, Wagner.911@osu.edu

[2] Division for Science and Innovation Studies, Administrative Headquarters of the Max Planck Society, Hofgartenstr. 8, 80539 Munich, Germany. bornmann@gv.mpg.de

[3] Amsterdam School of Communication Research (ASCoR), University of Amsterdam, Kloveniersburgwal 48, 1012 CX Amsterdam, The Netherlands , loet@leydesdorff.net

## Introduction

China's remarkable gains in science over the past 25 years have been well documented (e.g., Jin and Rousseau, 2005a; Zhou and Leydesdorff, 2006; Shelton & Foland, 2009) but it is less well known that China and the United States have become each other's top collaborating country. Recent studies have focused on the rapid growth of Chinese scientific publications overall. Adams (2010) reported that China had become the second largest producer of scientific articles behind the United States. In 2011, the Royal Society—the national science academy of the UK—issued a report using Scopus data in which this organization noted that China would overtake the U.S. in terms of numbers of publications within two years (Clarke *et al.*, 2011; Plume, 2011). Using Web of Science (WoS) data, Shelton & Leydesdorff (2012) found that the exponential growth of Chinese publications had slowed down to linear growth rates during the 2000s, and that China would not outpace the USA until a date beyond 2020 (see endnote 1). This article focuses on recent developments in the growth of collaboration between China and the United States, and examines in what fields this growth has occurred.

## Background of the China-USA science relationship

Science and technology has been a primary vehicle for growing the bilateral relationship between China and the United States since the opening of relations between the two countries in the late 1970s. From 1986-1997, China-USA cooperation made up 2.5 percent of U.S. internationally coauthored papers (NSB 2000 Table 6-61). By 2008, China's share of U.S. coauthored papers was up to 10.4 percent (NSB 2012 Table 5-19). By 2010, China had risen to second place (behind the UK) on the list of countries coauthoring with the United States (NSB 2012 Table 5-20). By 2011, China had become the top collaborating country with United States (Bornmann, Wagner, and Leydesdorff, under review).

At the official government-to-government level, the Joint Commission on S&T Cooperation oversees the bilateral scientific relationship. The Commission is an independent entity with no direct budget authority. The two governments have signed numerous high-level official agreements that provide a framework for cooperation (Sun, 2012; Suttmeier, 2010)[4]. The agreements have removed barriers and they provide platforms for cooperation, but they do not source actual research and development (R&D) projects, since they are only loosely tied to funding. This is particularly true on the U.S. side, where high-level S&T agreements (with any country) rarely commit the government to allocate funding. U.S. government R&D funding is generally awarded through a competitive process, with project funds being awarded to a U.S.-based entity. Oftentimes, a U.S. awardee has only a limited ability to share resources with a foreign partner. Thus, it is difficult to assess the extent or subjects of cooperation by examining political agreements. The publication record offers better insights into the subjects of cooperation and the locations of collaborators.

Both governments have funded some scientific cooperation with the other nation through direct and indirect channels. The U.S. government has supported China-U.S. cooperation through small amounts of direct funds for joint projects (e.g., U.S.-China Clean Energy Cooperation), funded outside of the R&D budgeting process. Other means of support for joint R&D are indirect and do not favor China over other countries. These sources include funds awarded to U.S. researchers who then cooperate with Chinese counterparts[5], (where the Chinese researcher is separately funded); as part of the U.S. government commitment to international cooperative projects such as the Human Genome Project or the International Thermonuclear Experimental Reactor; and in-kind contributions to global monitoring activities such as the earthquake research, ocean drilling, and weather monitoring.

Similarly, the Chinese government has supported enhanced cooperation with the United States with direct funding (of an unknown amount) as well as support for Chinese scholars who cooperated with the USA. The Chinese government has spent a significant amount of public funds to provide opportunities for Chinese scientists to travel to the United States to study or work in cooperation with U.S. counterparts. A number of studies and reports have documented the spectacular growth in the number of Chinese students traveling to the USA to study science, engineering, and computer science (see, for example, IIE, 2013). Relationships forged during education often continue after the student has returned to China, contributing to the pool of people coauthoring articles.

---

[4] A U.S. Department of State report, "U.S.-China S&T Cooperation," issued in July 2012, reports by agency on cooperative activities with China.

[5] The U.S. Department of State report notes that NSF awarded more than $15 million to U.S. researchers engaged in collaboration with China; these funds were competitively awarded, not set aside to fund research with China.

Suttmeier (2010) and Chen (2011) have documented the official government statements about bilateral cooperation. They describe the official approach at building bilateral cooperation where foreign delegations visited each other's countries to tour facilities and discuss exchanges of technical information. This paper reports on China-USA cooperation measured as published papers (articles, reviews, and letters) in scientific journals drawn from the Web of Science (WoS, Thomson Reuters). Published papers in scientific journals enable analysis of the results of scientific cooperation, irrespective of political relationships or agreements—or even of government spending. A review of published work abstracted in the WoS provides a mechanism to reveal the growth of scientific collaboration between the two countries. The publication record is expected to reveal the scientific links established by scientists themselves (rather than those listed as subjects in government-to-government agreements). We expected to find that scientific cooperation increases as the Chinese system transitions towards merit-based R&D and Chinese science improves in quality.

**Methodology and Data**

Data were derived in January 2014 from the WoS. For the years 2000 and 2012, China's gross output and the subset of papers coauthored with the United States were collected. Data for the years 2000 and 2012 are shown in Table 1. These data are limited to the journals listed in the WoS, which catalogued 161 international science journals published in China in 2012 (this is about 1.6 percent of the journals listed in WoS). The papers are parsed by fields of science that serve as the subjects of cooperation. Mapping was conducted using Google Maps. Data were also drawn from other literature, where appropriate.

The data were recreated in March 2014 from an analytical version of the Web-of-Science (WoS) at the Max Planck Society (Munich, Germany) to examine papers in the top 1% and 10% of most highly cited papers normalized against the corresponding subject categories, document types, and publication years. This second data collection was made using the Max Planck Digital Library (MPDL, Munich) which combines sets of the Science Citation Index Expanded (SCI-E), the Social Science Citation Index (SSCI), and the Arts & Humanities Index (A&HCI). These data were written up in an accompanying paper: Leydesdorff, Wagner, and Bornmann (2014; in preparation).

|  | *2000* | *2012* |
|---|---|---|
| Coauthored records | 2,594 | 20,371 |
| Addresses | 12,103 | 156,950 |
| Authors | 39,461 | 836,438 |
| Chinese addresses | 3,741 | 37,701 |
| US addresses | 4,756 | 51,183 |
| **Chinese first authorships** | **1,179** | **11,319** |
| **American first authorships** | **1,155** | **7,391** |
| Cities included in the mapping | 341 (≥ 5 times) | 370 (≥ 100 times) |

Table 1. Coauthored records of China-USA (articles, reviews, and letters from 2000 and 2012)

**Findings**

In the 35+ years since the two countries established a joint S&T agreement, coauthorships between China and the United States have increased from only 3 joint publications in 1978 to 20,371 in 2012 (Table 1). The growth on China's side has been part of an historically unprecedented emergence of a nation into world science. The addresses of practitioners working together rises from 12,000 to more than 150,000 institutional addresses in 2012 (Table 1). Both countries show rapid growth in the number of institutions working with the other nation. Chinese first authorships are most notable of the changes, from parity with the United States in 2000 to a third-more first authorships than the U.S. in 2012 (Table 1 and Figure 3).

Using Web of Science data, Figure 1 shows the growth and projected growth of publications as a percentage of world shares of publications from China (circles blue line), the USA (diamonds), and EU28 (boxes). At the rate of growth projected in Figure 1, China will outpace the United States in number of scientific publications around 2025. This trend line should be tempered with recent analysis by Zhou (2014) that shows that Chinese international publications dropped significantly in 2010, with a slight gain from there in 2011, but at a lower growth rate. Figure 1 shows the declining world shares attributed to the USA in the WoS, and the rise in world shares attributed to China. The line at the bottom of the figure (dark circles) shows the steady rise of China-USA coauthored papers as a percentage of world share of publications.

A notable asymmetry follows from the trends visible in Figure 1: the growth for China of coauthorships with the USA are part of China's overall growth pattern. China is increasing its output of papers in cooperation with many other countries during this period (Leydesdorff, Wagner, & Bornmann, under review).  At the same time, the USA does not experience growth of shares of coauthorships with other countries. In fact, the U.S. coauthorships with Russia, Israel, and Japan drops during this period, according to the National Science Board (2012). The rate of increase of collaboration between the U.S. with China may be unprecedented, outpacing over 35 years the rapid increases seen with eastern European countries in the 1990s following the breakup of the Soviet Union (Wagner, 2004).

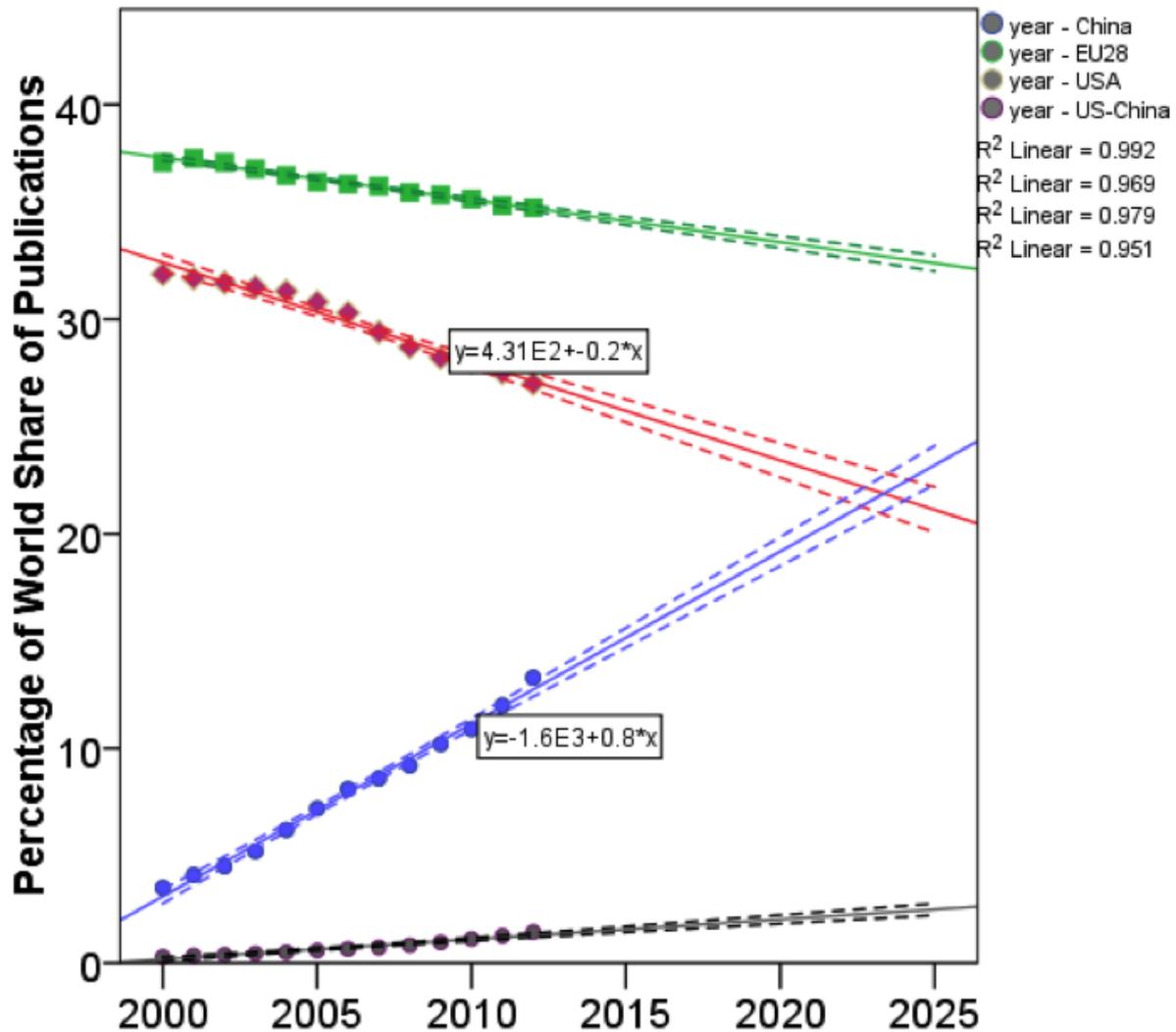

Figure 1. Percentages of world shares of publications from China, the USA and the EU28, and, co-authorships between China and the USA. Source: Leydesdorff, Wagner, and Bornmann (2014; under review).

At the bi-national level, China and the United States have increased coauthorships of scientific papers to the point where, in 2011, China became the United States' largest partner. (The USA has been China's top partner since 1999, according to Adams et al., 2009.) This corroborates Tang and Shapira (2011) who report that with more than two-fifths of all of China's internationally coauthored papers drawn from the WoS in 2008 showed U.S. collaborators. Moreover, Tang and Shapira (2011) report that about one-tenth of U.S. internationally coauthored papers involved Chinese co-authors in 2008. China's coauthorships with the USA rose from about 8% of the USA's international publications in 2000 to 11% by 2012, shown in Figure 2 (See end-note 1 for additional discussion). Of the coauthored papers identified for 2012, 65% are bi-national.

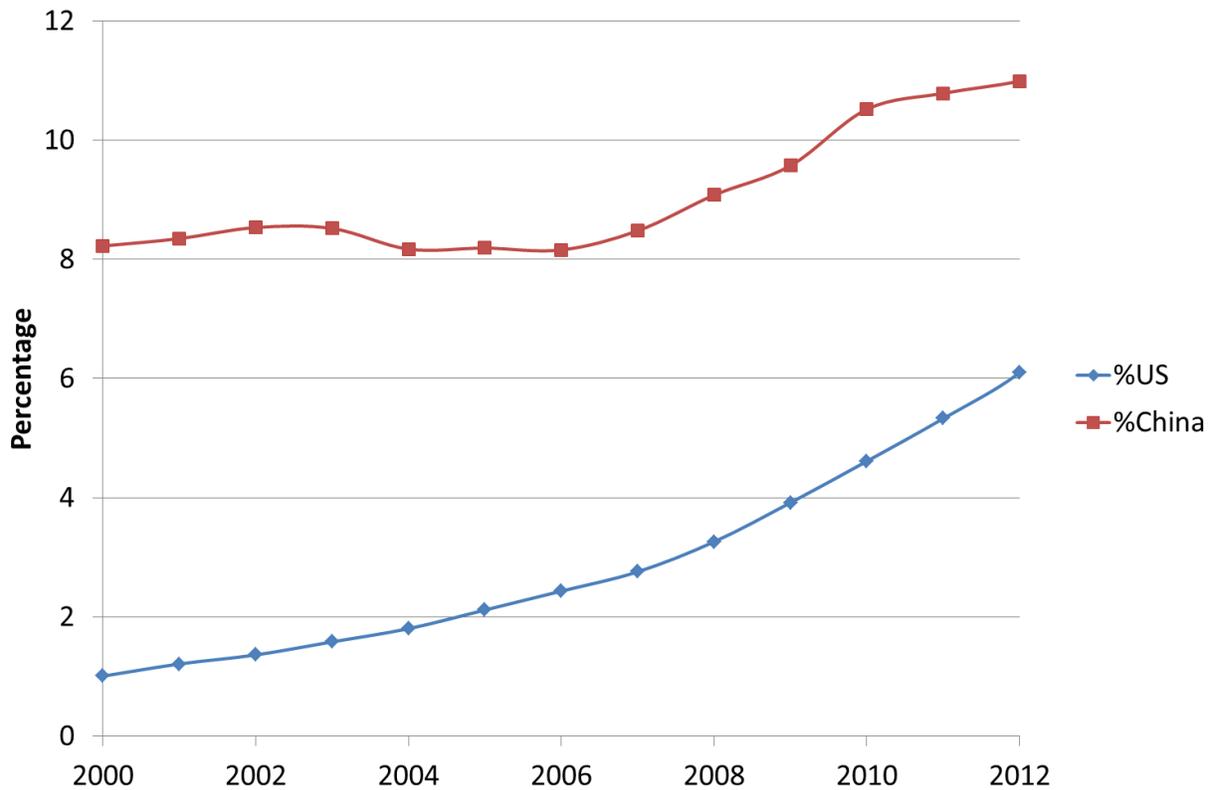

Figure 2: Co-authored documents between China and the USA as a percentage of the total numbers of papers of China and the USA, respectively.

Another significant change in the bi-national science relationship is seen in Figure 3 which shows the rise of Chinese first authors on coauthored papers. In 2000, Chinese and American scientists were equally likely to be the first author on a joint paper. By 2012, this had changed notably, with Chinese authors listed as the first author on joint papers about one-third more often than U.S. authors. First authorship usually indicates that the person took the lead on conducting the research and writing the paper.

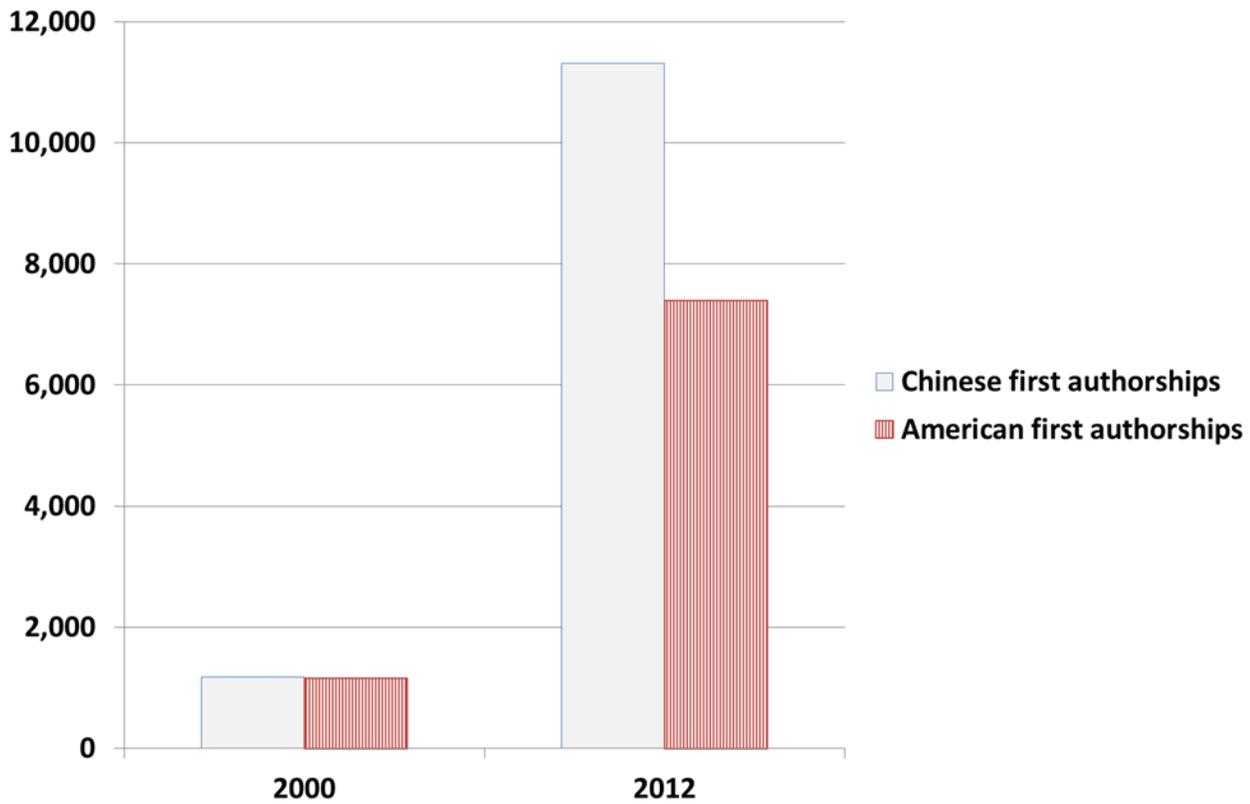

Figure 3. First authorships on China-U.S. coauthored papers, 2000 and 2012

In the earlier years of their bi-national relationship in science, China was more likely to co-publish with the USA in physics, chemistry, and basic materials sciences: fields of science linked more closely to heavy industry (Archambault, 2010). Kostoff (2009) showed that in 2002, the physical sciences dominated the subjects of China-USA cooperation. Over the 5 years between 2002 and 2007, China achieved parity with the USA in numbers of papers in the physical and engineering sciences (data drawn from the INSPEC database) (Kostoff, 2009). In coauthorship counts, our data show that fields of collaboration have remained relatively stable. Figure 4 shows the top 20 subjects of coauthored papers and the percentage share claimed by these papers of the 20,371 count for 2012. Among the subjects listed, three of them are new to top 50 subjects: *nanoscience-nanotechnology; cell biology; and neurosciences*. The figure shows a shift away from engineering-electrical-electronic, multidisciplinary physics, and applied mathematics (all of which were higher in 2002 than in 2012) and towards multidisciplinary sciences and chemistry, oncology, cell biology, and nanoscale sciences. The increase in these latter fields tracks with China's increased publication patterns of highly-cited articles in these fields (Fu & Ho, 2013).

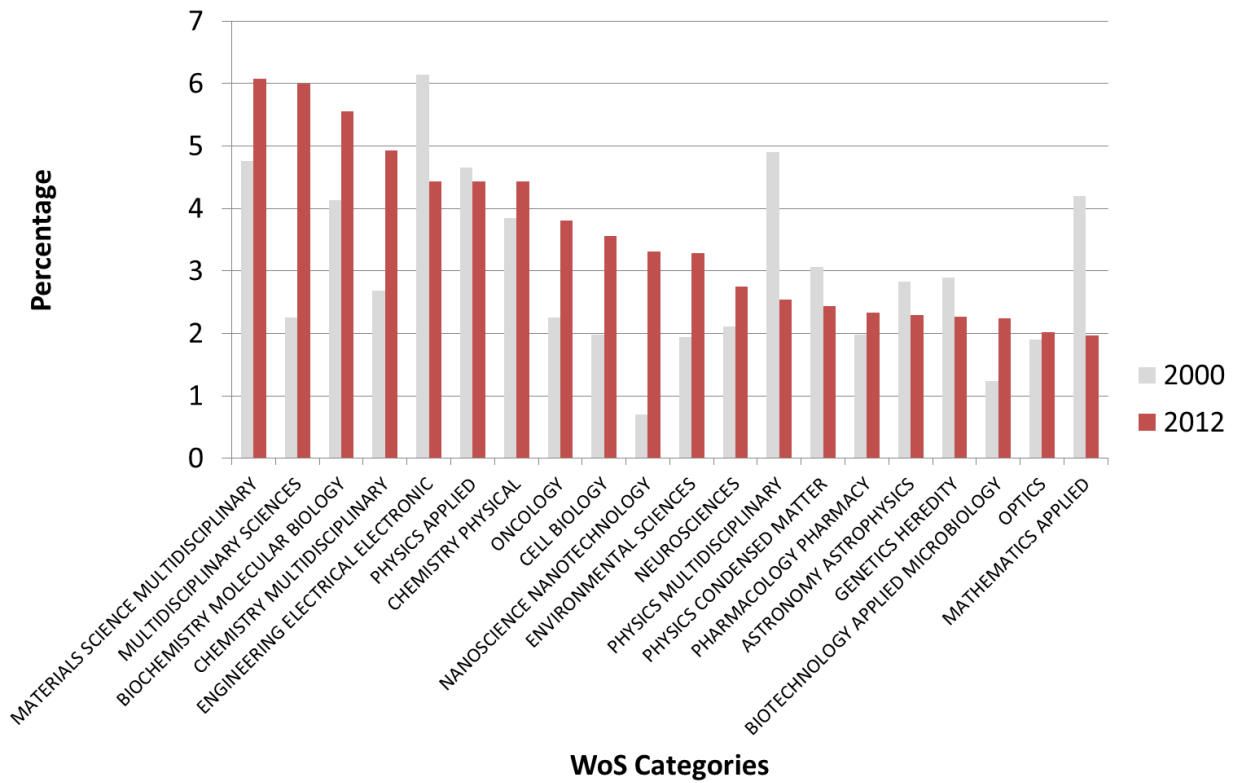

Figure 4. Numbers of records, and percentage of all articles by field of science (WoS categories) represented in China-USA coauthored papers, 2012. Source:WoS

Figure 5 shows a map of science in terms of WoS categories (Rafols, Porter, & Leydesdorff, 2010; cf. Klavans & Boyack, 2009) with nodes enlarged in proportion to the (logarithm of the) number of coauthored papers between China and the USA in those subjects. Materials science (black) stands out as a leading field (see also, Figure 4), which has remained a prominent subject for collaboration from the late 1990s and 2000s; the biomedical sciences (green) are prominent as a subject of collaboration, perhaps slightly more than has been the case in the past (Kostoff, 2009). Other sciences showing concentration are computer science (pink), environmental science and technology (orange), geosciences (brown), agricultural sciences (blue), and clinical medicine (red) (see also, Kostoff, 2009).

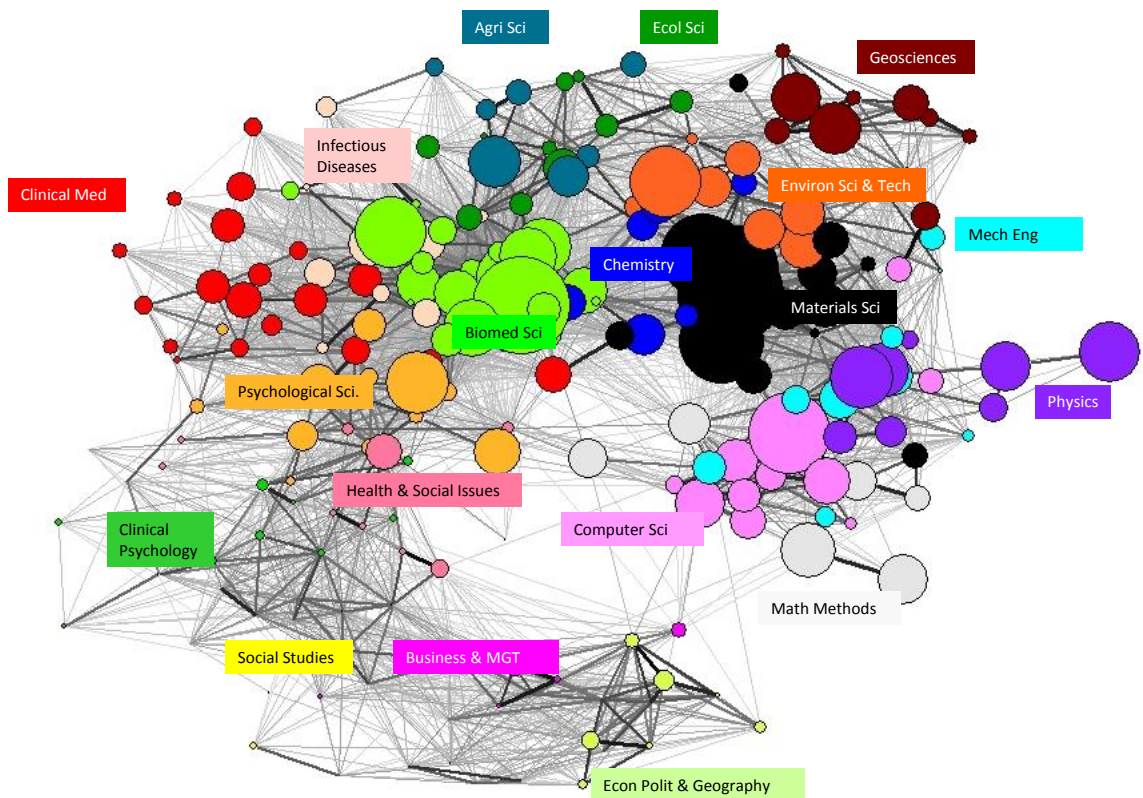

Figure 5. China-U.S. collaboration 2012 projected on the map of science in terms of WoS categories (Rafols et al., 2010).

The geographic concentration of Chinese coauthors working with U.S. counterparts shows growth in distribution and intensity over time. Figure 6 shows the 370 cities that are involved more than 100 times in internationally coauthored papers with at least one Chinese and American address. The node sizes are proportional to the logarithm of the number of papers, and the coloring corresponds to traffic lights: green as indication of more than expected in the top-10% highly-cited in this reference set (of 20,371 papers) and red for below expectation (based on z-testing these proportions; Bornmann & Leydesdorff, 2011). The figure shows that a large number of European (and other, e.g., Japanese) partners are also involved in the most green-colored participations.

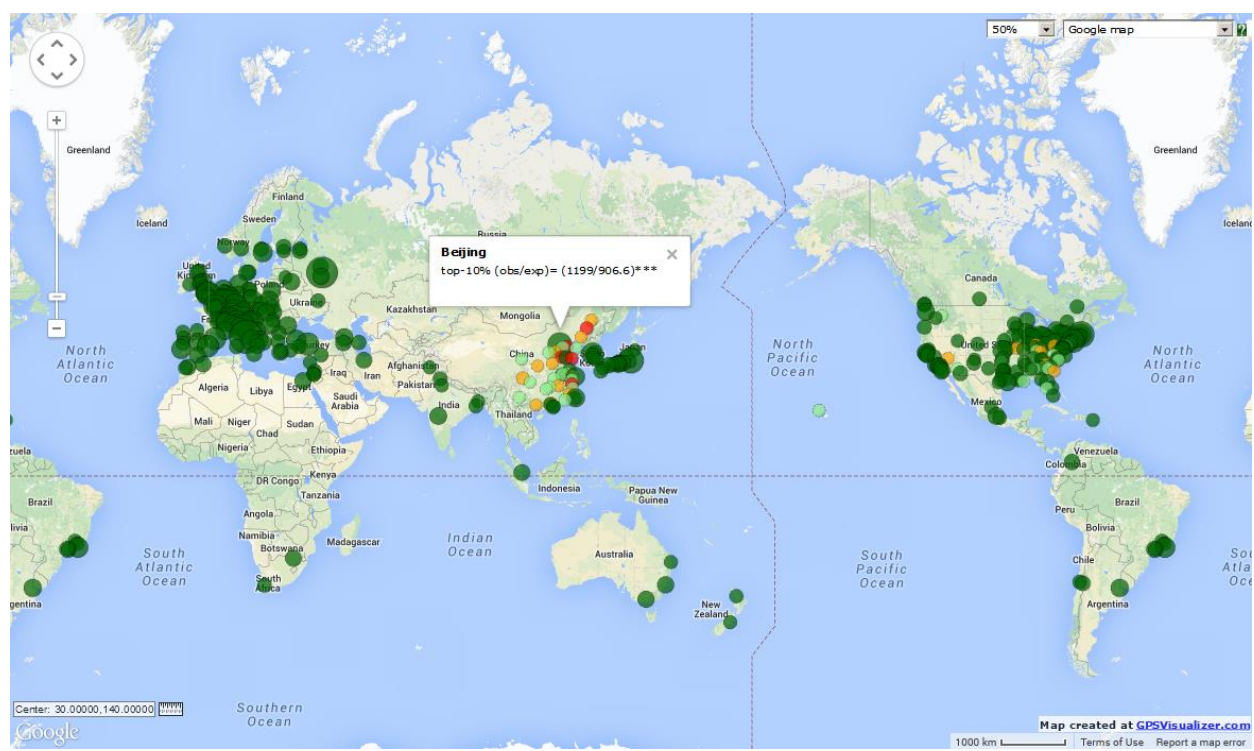

Figure 6: 370 city addresses involved more than 100 times in 20,371 papers internationally co-authored between China and the USA in 2012. An interactive version of this map is available at http://www.leydesdorff.net/china_us/ztest12.html .

A similar map for 2000 can be retrieved at http://www.leydesdorff.net/china_us/ztest00.html. This map shows the 341 cities worldwide that were involved more than five times in the 2,594 papers with at least one Chinese and one American address in 2000. The patterns have become more pronounced during the decade, but were equally global already in 2000.

China-USA coauthorships are cited above expectations. Leydesdorff, Wagner, and Bornmann (2014 Journal of Informetrics) show that China-USA coauthored papers appear in the top-10% and top-1% of cited papers at higher-then-expected rates. Network analysis of the collaborative patterns within the most-highly-cited articles for the year 2005, however, shows that China joins the group of highly-cited collaborators in terms of both binational and multinational collaborations. This network is very globalized.

**Discussion and Conclusion**

During the 2000s, the scientific relationship between China and the United States--as measured in coauthored papers--showed significant growth. By 2011, China had risen to become the USA's top collaborating partner, outpacing the UK, Japan, and Germany—nations that are long-time partners of the U.S. in science. Chinese scientists claim first authorship much more frequently than U.S. counterparts by the end of the decade. Collaborations increased in favor of nanoscale science and biological and medical sciences. Coauthored articles appear among the most highly-cited articles in the Web of Science.

The sustained rate of increase of collaboration with one other country is unprecedented on the U.S. side. By comparison, following the break-up of the Soviet Union, in 1998 after 7 years of openness, Russia's coauthorships accounted for 3.9% of U.S. international coauthorship (NSB 2000, Table 6-61); this percentage share put Russia in the $10^{th}$ position among the U.S.'s top 10 collaborating countries. The percentage share of U.S. international articles with a Russian coauthor(due to breakup of the Soviet Union, Russia was not counted in the 1980s) grew to 3.9% in 1997; this could be said to outpace China in the first years of openness, since China accounted for only 2.5 percent of U.S. internationally coauthored papers after 20 years of openness. But,

Russian co-authors did not sustain and grow the relationship. In 2008, after nearly 18 years of openness, Russia's share of coauthorship with the U.S. had dropped to 2.7% of U.S. papers (NSB 2010), and it dropped again by 2012 to 2.4% (NSB 2014). Also, by point of contrast, the United States government spent more than $300 million per year (in then-year dollars) during the 1990s to encourage civilian uses of Russian science and to build collaborative relationships (Wagner et al., 2002)—where no counterpart of expenditure is committed to building a relationship with China.

From a modest beginning, China continued to grow its share of U.S. international coauthorships, even in the absence of targeted funds (on the U.S. side) such as those provided to the U.S.-Russia relationship. This suggests that the scientific relationships being forged were emerging more from the interests of scientists to link to one another than from political incentives (as might have been an influence on the U.S.-Russia relationship in the first decade after opening). Moreover, it also suggests that China continued to improve its domestic scientific capacity over the past two decades, again in contrast to Russia, where bilateral scientific relationships with the U.S. have dropped as Russia has dropped in scientific quality measures.[6]

It remains to be seen if political opposition expressed in U.S. law by the Wolf Amendment (April 2011) influences cooperation with China (Forbes, 2011). However, as noted, U.S. official agreements or statements rarely commit government funding: the growth in cooperation appears to be driven bottom-up by the interests of scientists rather than top-down by governmental edict or other constructed opportunity provided by preferential funding, so political opposition may have little affect on actual scientist-to-scientist cooperation.

During the 2000s, the China-U.S. relationship grew within a global science system, as well. The global system has also become more collaborative overall (Wagner & Leydesdorff, forthcoming). Increased international coauthorships are part of an overall trend towards global linkages. Even within this global trend, the China-U.S. growth outpaced the growth of other relationships. The National Science Board shows that the increases in coauthorship relationships between the United States and many other countries were generally in the range of 1-2% increase, while with China, the rate tripled from 5% of coauthored articles in 2002 to 16.2% in 2012 (NSB 2014 Table 5-23, see endnote 1. (In the same decade significant drops in coauthorships are recorded with the U.S.: Japan, Russia, and Israel (NSB 2014 Table 5-23)).

---

[6] Russia's ranking in the SCImago Journal and Country Rank also dropped through the decade from 9th in the world in 2000 to 16th in the world in 2012. www.scimagojr.com.

A number of reasons may explain this sustained rise in scientific partnership between China and the United States, including: 1) increased R&D spending on China's part---increasing resources available to cooperate, 2) increased number of internationally educated Chinese scientists and engineers—increasing the number of people available to work abroad and/or co-author internationally, and 3) changes to the databases that count China's contributions to science. Each is discussed briefly below.

1)   Certainly we know that China has greatly increased its investments in science and technology spending (Suttmeier, 2010; Shelton and Foland, 2009), which contributes to its domestic capacity to collaborate. The Chinese government has increased R&D spending (GERD)[7] up to the level of the EU27 according to the OECD (2008). Internal Chinese government policy has encouraged private R&D spending, all with a view towards supporting economic growth (Cao, Suttmeier, & Simon, 2006). According to the NSB, Chinese spending on R&D has increased by 18% annually (adjusted for inflation) (NSB, 2014); the NSB (2014) also reports that Chinese R&D spending rose to be 15% of global R&D spending in 2011. Intramural spending on R&D within industry more than doubled during the decade (UNESCO, 2010).

2)   The number of researchers working in China has increased significantly during the decade from 695,000 in 2000 to 1,600,000 in 2008, according to UNESCO (2010). Chinese universities have increased the rate of growth of human capital, partly demonstrated in their increased publishing in WoS-listed journals. Financial incentives are offered within China to publish in refereed journals. The Chinese government has sponsored workshops to teach scientists how to write for and publish in English-language journals (Cargill & O'Connor, 2006). We know that those publishing in WoS journals gains international visibility, and scientists working at world-class levels are more likely to coauthor (Narin et al., 1991). Also, the quality of Chinese science appears to be improving (Kostoff 2009; Jin & Rousseau, 2005b), possibly adding to their attractiveness to U.S. counterparts. Chinese scientists who have been living and studying in the U.S. have been encouraged to return to China, and it can be surmised that they are likely to co-publish at the global level. Chinese students studying in the USA and returning home may be retaining ties and co-publishing with former teachers and colleagues. These changes have increased the number of Chinese scientists who seek to co-publish in archival journals.

---

[7] GERD is gross expenditure on research and development.

3)  Finally, the indexing services that abstract scientific publications have increased coverage of Chinese science. The two primary indexing services, Scopus (Elsevier) and WoS databases, have sought to include a growing number of Chinese journals. WoS now lists 161 peer-reviewed journals published in China. (This number is still a very small percentage of all journals published in China, which number more than 6500 according to the China Scientific and Technical Papers and Citations Database.) The addition of new journals in WoS is an attempt to address the English language bias within science (Van Leeuwen et al., 2000) although we can assume that some language bias and possibly a cultural bias still exists. (In fact, papers with a U.S. address continue to be more likely to cite papers from Europe; similarly, Chinese papers are more likely to cite South Korea and Taiwan.)

Chinese scientists are actively seeking to partner with counterparts in the United States. Some see this targeted outreach on China's part to cooperate with the United States as a threat to U.S. dominance in science (Freeman, 2006). It can also be seen as an opportunity to tap fresh talent and ideas. As the West faces a period of budget austerity, science and technology projects funded in China may help fill some gaps if the activities are accessed with the goal of gaining efficiencies and disseminating know-how. A policy shift may be needed to take full advantage of these opportunities. U.S. government agencies, for example, may need to shift investment towards tracking and monitoring science in China and making it available to U.S.-based scientists.

Chinese science faces challenges of its own that potentially hinder future growth and collaborations. The most pressing challenge is addressing continuing charges of fraud and plagiarism in a way that builds trust in collaborative opportunities. Lin Songquin, a member of the Chinese Academy of Sciences, expressed concern about academic corruption in Chinese science saying that a midguided academic evaluation system creates conditions for fraud and plagiarism (Lin, 2013). Suttmeier, Cao and Simon (2006) point to a possible "technology trap" caused by China's over-reliance on imported technology, where knowledge capacity is not embedded locally in a way that can take up the technologies being imported from abroad. Similarly, Wilsdon and Keeley (2011) pointed to structural barriers to openness within the Chinese society that affect the ability of scientific inquiry to grow.

Our data support those of Braun & Dióspatoni (2005) that the dominance of the USA in terms of editorial control of journals is still evidenced by results (Braun *et al.*, 2007). Many thousands of Chinese journals are not listed in the WoS, although this is the case for many U.S. scientific publications as well (Wagner & Wong, 2011), and the extent of bias is diminishing. The data show that the scientific system as a whole is growing, and new members from formerly absent countries are contributing to the pool of knowledge.  As they do, the system as a whole benefits from new knowledge (Wagner, 2008).  From this perspective, the USA and other scientifically-advanced countries are gaining new colleagues and partners as well as access to new resources as other countries develop their scientific capacities.

On the other hand, it is clear that the location of scientific research contributes to the capacity to create, absorb, and retain knowledge as well as to create innovative products (Acs et al., 2006). The data presented here  suggests that China's economy is significantly enhancing innovative capacity.  To the extent that collaboration creates access to these knowledge sources, China will certainly benefit by the growth of knowledge gained in part by collaboration with U.S. counterparts. To the extent that local-learning-loops feed the efficiency of Chinese manufacturing and invention, the growth of China's scientific capacity may contribute to enhanced industrial competitiveness in higher value-added products. A similar benefit on the U.S. side would require a deliberate strategy within U.S. government agencies to identify and link to important research in China (Freeman 2006).

-------------------------------------------------

**Endnote 1**

The United States government tracks indicators of international activities in science and engineering, and publishes these data in a biennial report issued by the National Science Board through the auspices of the National Science Foundation. The Science & Engineering Indicators report uses a subset of the Science Citation Index and the Social Sciences Citation Index (Web of Science) data, thus, their results are not directly consonant with the results herein, which draw from a broader dataset. The Science & Engineering Indicators 2014 report indicated that the United States share of Chinese international articles was close to 48% in 2012. China's share of U.S. international articles is reported to be 16.2% in 2012, where the calculation made by the authors using WoS data showed a lower percentage share. See NSB 2014, Table 5-23. The choice of database from which the data are drawn influences the results in that the Web of Science lists fewer journals overall than Scopus, for example. The reports noted by Archambault (2010) and the Royal Society (2011) used Scopus data. INSPEC is another database that reports on engineering and physical sciences; this database was used by Kostoff (2008).